\documentclass[%
 preprint,
 showkeys,
 amsmath,amssymb,
 aps, physrev,
]{revtex4-2}

\usepackage{graphicx}
\usepackage{dcolumn}
\usepackage{bm}
\usepackage{xcolor}
\usepackage[normalem]{ulem}

\begin{document}
\title{\textbf{Change in the Order of a Phase Transition in the 2D Potts Model with Equivalent Neighbours.} 
}%

\author{Petro Sarkanych}
 \email{Contact author: petrosark@gmail.com}
\affiliation{
 Yukhnovskii Institute for Condensed Matter Physics of the National Academy of Sciences of Ukraine, Lviv, 79011, Ukraine;
}%
\affiliation{
$\mathbb{L}^4$ Collaboration and Doctoral College for the Statistical Physics of Complex Systems, Lviv-Leipzig-Lorraine-Coventry, Europe
}

\date{\today}

\begin{abstract}
Two dimensional Potts model is a classical example when the symmetry of the order parameter controls the order of a phase transition: on a square lattice with nearest-neighbours interaction, when the number of states $q$ is less than or equal to 4, the second-order phase transition is observed, while for $q>4$ the first-order phase transition occurs. Recent research shows that even when the number of states is fixed, increasing the interaction range allows one to reach the point where the order of the phase transition changes. We focus on a $q=3$ 2D Potts model and, from the analysis of the partition function zeros, locate the number of interacting neighbours that change the order of the phase transition.
\end{abstract}

\maketitle


\section{Introduction}

In statistical physics, the universality hypothesis tells that the critical properties of a system near the second-order phase transition point depend on the dimensionality, symmetry of the order parameter and the interaction range \cite{universality}. 
For example, the Ising model undergoes a second-order phase transition both in two and three dimensions, but the critical exponents are completely different \cite{ising2017fate}. 
Or the $O(n)$ model where the $n$ defines the number of components in the order parameter and, therefore, affects the critical behaviour \cite{stanley1968dependence}. 
When discussing interaction ranges, a distinction is made between short-range and long-range interacting systems. For the standard Ising model on a square lattice, it is well established that only in the limit where the interaction radius becomes infinite do the critical properties shift from those characteristic of the nearest-neighbour case to those of the mean-field solution \cite{luijten1996medium}. When long-range interactions are introduced in a tunable fashion, however, the critical exponents become dependent on the specific details of the interactions \cite{cannas1995one, sarkanych2015phase}.

The possibility of a finite-range interaction to change the order of a phase transition was earlier investigated in Refs. \cite{Gobron2007,Biskup2003}. 
It was rigorously proven that even on a square lattice $q=3-$states Potts model can change the order of a phase transition when the interaction radius is large enough but still finite \cite{Gobron2007}.

The Hamiltonian of the Potts model with finite interaction range is the following
\begin{equation}
\label{hamiltonian}
    H=-\cfrac{1}{z}\sum_{\{ij\}}\delta_{s_i,s_j},
\end{equation}
where $s_i=1,\ldots,q$ is the spin variable and $z$ is the number of interacting neighbours.
The sum goes over all interacting spins. The normalisation factor $\cfrac1z$ ensures that the mean-field limit $z\to\infty$ is properly defined.

This model has already been investigated with a help of Monte-Carlo simulations to find the number of interacting neighbours $z$ that are needed to change the order of the phase transition \cite{Bloete2016}. 
The Authors performed simulations using the Luijten-Bl\"ote algorithm and analysed the scaling of the Binder cumulant.  
They concluded that $z=80$ is a marginal value that separates the regions with different critical regimes.
For $z < 80$, the phase transition is continuous and belongs to the same universality class as the nearest-neighbour case. In contrast, for $z > 80$, the system undergoes a first-order phase transition, consistent with predictions for the $d = 2, q = 3$ Potts model with infinite-range (mean-field) interactions.

In this paper, we aim to provide a more precise estimate of the critical $z$-value that separates these two regimes of differing critical behaviour. To achieve this, we employ an analysis based on the partition function zeros, a method known to yield accurate estimates of critical exponents even with limited data~\cite{Moueddene2024a, Moueddene2024b, Sarkanych2021}.
Furthermore, we relax the constraint previously imposed on  $z$ —that only values corresponding to complete coordination spheres are considered~\cite{Bloete2016}—and instead allow $z$ to vary in steps of 4, preserving the lattice symmetry.

The rest of the paper is organised as follows: in Section \ref{II}, the simulations algorithm and the methods to analyse partition function zeros are described.
In Section \ref{III}, simulation results are provided and obtained values of critical exponents are given.
Our results are reviewed in the Discussion section.

\section{Method} \label{II}
\subsection{Algorithm}
For the purposes of our study, we have employed the Fukui-Todo algorithm~\cite{FukuiTodo}. 
This is a cluster Monte-Carlo algorithm that performs in $O(N)$ time even for the systems with long-range interactions.
It builds upon the well-known Fortuin-Kasteleyn representation of the partition function~\cite{FK}, where each bond  $l$ in the system is characterised not only by its strength $J_l$, but also by an associated bond variable $k_l = 0, 1$. 
A value of $k_l = 0$ corresponds to an inactive (or absent) bond, while $k_l = 1$ indicates an active (or present) bond. 
The partition function in this representation is calculated as
\begin{equation}
\label{zfk}
    Z_{FK}=\sum_{\{s\}} \sum_{\{g\}} 
\prod_{l=1}^{N_b} 
\left( (e^{\beta J_l} - 1)^{k_l} \left[ 1 - k_l (1 - \delta_{s_i s_j}) \right] \right),
\end{equation}
where the first sum runs over all possible spin configurations and the second one over all bond configurations; $N_b$ is the number of bonds; $\beta$ is the inverse temperature; and $\delta$ is Kronecker's delta. 
Summing over all possible bond configurations in Eq. (\ref{zfk}) recovers a usual formula for the Gibbs partition function
\begin{equation}
\label{Gibbs_partfunc}
    Z_{G}=\sum_{\{s\}} e^{-\beta H}. 
\end{equation}

For a variety of cluster Monte-Carlo algorithms, the sampling process can be split into two steps: graph generation and cluster flip. Two illustrative examples are Swendsen-Wang \cite{swendsen1987nonuniversal} and Luijten-Bl\"ote \cite{luijten1995monte} algorithms.
The first step in Swendsen-Wang approach  is done by scanning all the bonds and activating them with a probability
\begin{equation}
    p_{l} = \delta_{s_i,s_j}\left[ 1 - \exp(-2 \beta J_{l}) \right],
\end{equation}
while the formed clusters are then flipped with a probability $1/2$. 
Scanning all the bonds for the system with the long-range interaction takes $O(N^2)$ time, making this approach rather slow.

In the  Luijten and Bl\"ote approach, the activation process is split into two parts: selecting a candidate bond with a probability 
\begin{equation}
    p_{l} = 1 - \exp(-2 \beta J_{l}),
\end{equation}
and subsequently activating it if this bond connects two spins in the same state. 
This allows to circumvent the scanning of all the nodes and, with the help of a binary search-based technique, to reduce the time and achieve $O(N\rm{log}N)$ performance.

The key extension introduced by the Fukui-Todo algorithm is that the bond variable $k_l$ is allowed to take any non-negative integer value, $k_l = 0, 1, 2, \ldots$, with values drawn from a Poisson distribution:
\begin{equation}
\label{poisson}
    f(k_l; \lambda_l) = \frac{e^{-\lambda_l} \lambda_l^{k_l}}{k_l!},
\end{equation}
where $\lambda_l = \beta J_l$.

As before, bonds with  $k_l = 0$ are treated as inactive, while those with $k_l > 0$ are considered active. The partition function in this representation takes the form:

\begin{equation}
\label{partfunc}
Z_{\rm FT} = \sum_{\{s\}} \prod_{\ell=1}^{N_{\rm b}} \sum_{k_\ell=0}^{\infty} \Delta (\sigma_\ell, k_\ell) V_\ell (k_\ell)
\end{equation}
where the first sum runs over all possible spin configurations, the product goes over all bonds $N_{\rm b}$, whereas the second sum is taken over all possible bond states and
\begin{align}
\sigma_l &= \left\{ 
\begin{array}{l}
1 \qquad \mbox{if the bond $l$ connects the spins in the same state} \\
0 \qquad \mbox{otherwise,}
\end{array}
\right. \\
\Delta(\sigma_l, k_l) &= \left\{ 
\begin{array}{l}
0 \qquad \mbox{if $k_l\ge 1$ and $\sigma_l=0$} \\
1 \qquad \mbox{otherwise,}
\end{array}
\right. \\
V_l(k_l) &= \frac{ (\beta J_l)^{k_l}}{k_l!}.
\end{align}
Performing summation over $k_l$ in Eq. (\ref{partfunc}) one shows that this definition of the partition function in the extended phase space is identical to a conventional Gibbs definition (\ref{Gibbs_partfunc})

With this extension of the phase space the Fukui-Todo algorithm can exploit the fact that Poisson process describes independent events, therefore instead of generating a variable for each bond, one can generate a single random number from a Poisson distribution with the mean $\lambda_{\rm tot}=\beta\sum_l J_l$ and then distribute this value among the bonds. 
Generating a random number from Poisson distribution takes $O(\lambda_{\rm tot})$ time, which is $O(N)$ for models where the ergodicity condition holds. 
The distribution step can be performed in $O(1)$  using, for example, the Walker method of alias \cite{walker1977efficient}\footnote{This require to have a precomputed weights of each bond, which can be done once for each set of model parameters and remains constant throughout the run.}. 
This allows to do a Monte-Carlo update in the Fukui-Todo approach in $O(N)$ time. 
Although, in this algorithm only $K=\sum_l k_l$ is directly measured, internal energy and specific heat can be easily calculated:
\begin{equation}
	E = - \frac{\partial}{\partial \beta} \ln \Big[ \sum_{\{s\}} \prod_{\ell=1}^{N_{\rm b}} \sum_{k_\ell=0}^{\infty} \Delta (\sigma_\ell, k_\ell) V_\ell (k_\ell) \Big] 
	= J - \frac{1}{\beta} \Big\langle K \Big\rangle,
	\end{equation}
	\begin{equation}
    \label{cheat}
	C = - \frac{\beta^2}{N} \frac{d E}{d \beta} = \frac{1}{N} \left[ \Big\langle K^2 \Big\rangle - \Big\langle K \Big\rangle^2 - \Big\langle K \Big\rangle \right],
	\end{equation}
where $J = \sum_l J_l$. Here the angular brackets denote thermodynamic averaging with the partition function given by Eq. (\ref{partfunc}). 
In equation (\ref{cheat}) for the specific heat, the variable $K$ plays a role analogous to that of the energy. However, in addition to the variance of $K$, there appears an extra term, $\langle K \rangle$. 
This distinction becomes particularly relevant for systems with long-range interactions, where evaluating the energy requires summation over all interacting pairs of spins, resulting in a computational complexity of $O(N^2)$. 
For the model under consideration, the energy evaluation instead scales as $O(Nz)$, which nevertheless presents a computational challenge for sufficiently large interaction ranges.
Moreover, the magnetisation $m$ can be evaluated at each iteration of the algorithm without a significant impact on performance, since only the states of individual spins need to be accessed. 
The simultaneous availability of both the specific heat and the magnetisation facilitates the determination of the critical exponents and the critical temperature through established finite-size scaling (FSS) analysis techniques~\cite{Landau}.

Recently, the Fukui-Todo algorithm was extended to enable the partition function zeros analysis \cite{Sarkanych2021}. In the following subsection we provide the main ideas of this extension and how the partition function zeros can be analysed.

\subsection{Partition function zeros}
Analysis of the partition function zeros in a complex plane has been shown to be an excellent tool to analyse the critical properties of a system \cite{Moueddene2024a,Moueddene2024b,Sarkanych2021}. 
Not only this approach allows for precise estimates of critical exponents, but also it can do so using simulation data of smaller system sizes \cite{Honchar2024,Honchar2025,Moueddene2024b}. 
In this section we will describe how we will use the partition function zeros in the complex temperature plane - the Fisher zeros \cite{Fisher} to understand the properties of a phase transition in the $q=3$ 2D Potts model with equivalent neighbours.

In the conventional Monte-Carlo simulations, when the energy is measured at each step, the Fisher zeros coordinates can be found from the condition
\begin{equation}
    \sum_{n} e^{-\beta E_n}=0
\end{equation}
where the summation is taken over all measurements. Allowing complex values $\beta=\beta_{\rm re}+i \beta_{\rm im}$ this transforms to the system of equations
\begin{gather}
\label{explicit}
    \sum_{n}\cos(\beta_{\rm re}E_n)=0 \\
\nonumber \sum_{n}\sin(\beta_{\rm im}E_n)=0.
\end{gather}
As mentioned in the previous subsection, the only two quantities directly measured in the Fukui-Todo algorithm are $K$ and $m$. 
For systems with long-range interactions, computing the energy requires $O(N^2)$ time, which significantly degrades the performance of the Fukui-Todo update. 
In the case of the $q = 3$ 2D Potts model with $z$ equivalent neighbours, energy computation requires $O(zN)$ time. 
Although this is an improvement, it is still considerably slower than the update step alone. 
Therefore, the use of Eqs.~(\ref{explicit}) becomes impractical in this context.

To circumvent this issue, a new approach was proposed in Ref.~\cite{Sarkanych2021}. 
It is based on the reweighting method, which was adapted for Fukui-Todo simulations in Ref.~\cite{Flores2017}. 
If the simulations are performed at an inverse temperature $\beta$, the following relation holds:
\begin{equation}
    \label{relation}
    Z_{\rm FT}(\beta')=Z_{\rm FT}(\beta)\Big\langle\left(\frac{\beta'}{\beta}\right)^K\Big\rangle_\beta \quad,
\end{equation}
where it is explicitly noted that the averaging is performed at fixed temperature $\beta$. 
This allows to search for the partition function zeros $\beta'\in\mathbb{C}$ using the equation
\begin{equation}
    \label{condition_fisher1}
    \sum_{n}\left(\frac{\beta'}{\beta}\right)^{K_n}=0,
\end{equation}
where the summation is taken over all measurements.

When analysing the simulation data, it is also useful to exclude the lowest $K$ value in the dataset $K_{\rm min}$ as this procedure transforms the last equation to 
\begin{equation}
    \label{condition_fisher3}
    \sum_{n}\left(\frac{\beta'}{\beta}\right)^{K_n-K_{\rm min}}=0,
\end{equation}
and allows for faster and more accurate zeros coordinates finding.

With the Fisher zeros coordinates at hand, there are two popular approaches to extract critical properties. The first approach is to analyse the density of zeros \cite{Janke2002}. 
Assuming that the $j-$th zero $\beta_j$ of a system sized $L$ is located on a distance $r_j(L)$ from the critical point, one can write the density of zeros as
\begin{equation}
   g_L(r)=L^{-d}\sum_{j}\delta(r-r_j(L)), 
\end{equation}
where the index $j$ show the sequential order of a Fisher zero and $\delta(r)$ is the Dirac $\delta-$function. 
The cumulative distribution function is then
\begin{equation}
    G_L(r)=\int_0^r g_L(s)ds.
\end{equation}
This function simply counts the number of partition function zeros that are located closer than $r$ from the critical point. 
When $r=r_j(L)$ one can define
\begin{equation}
    G_{L}(r_j)=\cfrac{2j-1}{2L^d}
    \label{density_zero},
\end{equation}
In case of the second-order phase transition, the cumulative zeros density scales according to the formula \cite{Janke2002}
\begin{equation}
    G_{L}(r_j)=\cfrac{2j-1}{2L^d}\propto r_j^{2-\alpha}
    \label{density_scaling},
\end{equation}
where $\alpha$ is the specific heat critical exponent. 
For the first-order phase transition, the cumulative zeros density scales as $G_{L}(r_j)\propto r_j$, which is effectively $\alpha=1$.
In case of the $q=3$ 2D Potts model, the expected values of $\alpha$ are given in Table \ref{tab0}.
For practical purposes $r_j$ is often replaced by the imaginary part of a zero ${\rm Im}\,z_j$ \cite{Moueddene2024a,Moueddene2024b,Honchar2024,Honchar2025,Sarkanych2021,sarkanych2015phase}. 
The main advantage of this approach is that it allows to use the coordinates of multiple zeros for the same system size and, therefore, reduces the computational cost.

The second approach is to look at the coordinates of the closest to the real axis zero $\beta_1$ as they scale with the system size according to the ansatz \cite{Itzykson} 
\begin{eqnarray}
\nonumber \mbox{Re}\, \beta_1=\beta_c+A\cdot L^{-1/\nu}\\
\label{ansatz}\mbox{Im}\, \beta_1=B\cdot L^{-1/\nu},
\end{eqnarray}
where $\beta_c$ is the inverse critical temperature and $\nu$ is the correlation length critical exponent for the second-order phase transition. 
For the first-order phase transition, the coordinates of the first zero scale proportionally to $L^{-d}$ where $d$ is the space dimensionality, or effectively $\nu=1/d$.
In Table \ref{tab0} we summarise the values of critical exponents for different critical behaviours. We will detect the change in the order of the phase transition via changes in critical exponents values.
For the nearest neighbours $q=3-$state Potts model on a square lattice one expects to get $\alpha=1/3$ and $\nu=5/6$ \cite{Wu1982}. 
While for the first-order phase transition observed on a complete graph with infinite interaction range $z=\infty$,  the effective values of critical exponents are $\alpha=1, \nu=1/2$. The values of critical exponents of the 2D Potts model in the tricritical regime are $\alpha=5/6$ and $\nu=7/12$ \cite{nienhuis1987coulomb}.

\begin{table}[h]
    \centering
\begin{small}
    \begin{tabular}{|c|c|c|c|c|}
        \hline
         & Scaling & \,Nearest neighbours, $z=4$\,& \,Tricritical point\,  & \,Mean-field, $z=\infty$ \,\\
        \hline
        $G(r)$ & $r^{2-\alpha}$ & $\alpha=1/3\approx0.333$ &  $\alpha=5/6\approx0.833$ & $\alpha=1$ \\
        \hline
        $\mbox{Im}\beta_{1}$ & $L^{-1/\nu}$ & $\nu=5/6\approx0.833$ & $\nu=7/12\approx0.583$ & $\nu=1/d=1/2$ \\
        \hline
    \end{tabular}    
    \end{small}
    \caption{Different scaling of the $q=3$ 2D Potts model depending on the critical regime. For the nearest neighbour case, this model undergoes a second-order phase transition with critical exponents $\alpha=1/3, \, \nu=5/6$ \cite{Wu1982}. In the mean-field case, the transition is of a first-order with effective values of critical exponents $\alpha=1,\, \nu=1/2$. For the tricritical regime, the expected values are   $\alpha=5/6,\, \nu=7/12$ \cite{nienhuis1987coulomb}.\label{tab0}}
\end{table}
\section{Results} \label{III}

We applied the approach described in the previous section for the  $q = 3$ 2D Potts model with equivalent neighbours~(\ref{hamiltonian}), considering the square lattice and progressively increasing the number of interacting neighbours $z$ to track changes in the order of the phase transition.

In Ref. \cite{Bloete2016} it was found that the order of the phase transition changes near the value $z_c=80$. 
Therefore, we performed simulations in the region $z=68-100$ near the critical temperature for the system sizes indicated in Table \ref{tab1}. 
The number of interacting neighbours was changed with the step $\Delta z=4$ so that the system's square lattice symmetry is unaffected. For each of the system sizes, we collected 400000 uncorrelated measurements.

\begin{table}[h]
    \centering
    \begin{small}
    \begin{tabular}{|c|c|c|c|c|c|c|c|c|c|}
        \hline
         & $z=68$ & $z=72$ & $z=76$ & $z=80$ & $z=84$ & $z=88$ & $z=92$ & $z=96$ & $z=100$ \\
        \hline
        $L_{\rm min}$ & 32 & 32 & 32 & 32 & 32 & 32 & 32 & 32 & 32\\
        \hline
        $L_{\rm max}$ & 432 & 368 & 352 & 352 & 320 & 288 & 256 & 256 & 288 \\
        \hline
    \end{tabular}    
    \end{small}
    \caption{Simulation data available for each interaction range $z$. $L_{min}$ indicates the minimal system size that was simulated and $L_{max}$ - is the maximal system size. For all the cases, the step in the system size is $\Delta L=16$.}
    \label{tab1}
\end{table}

\begin{figure}
    \centering
    \includegraphics[height=0.42\textwidth]{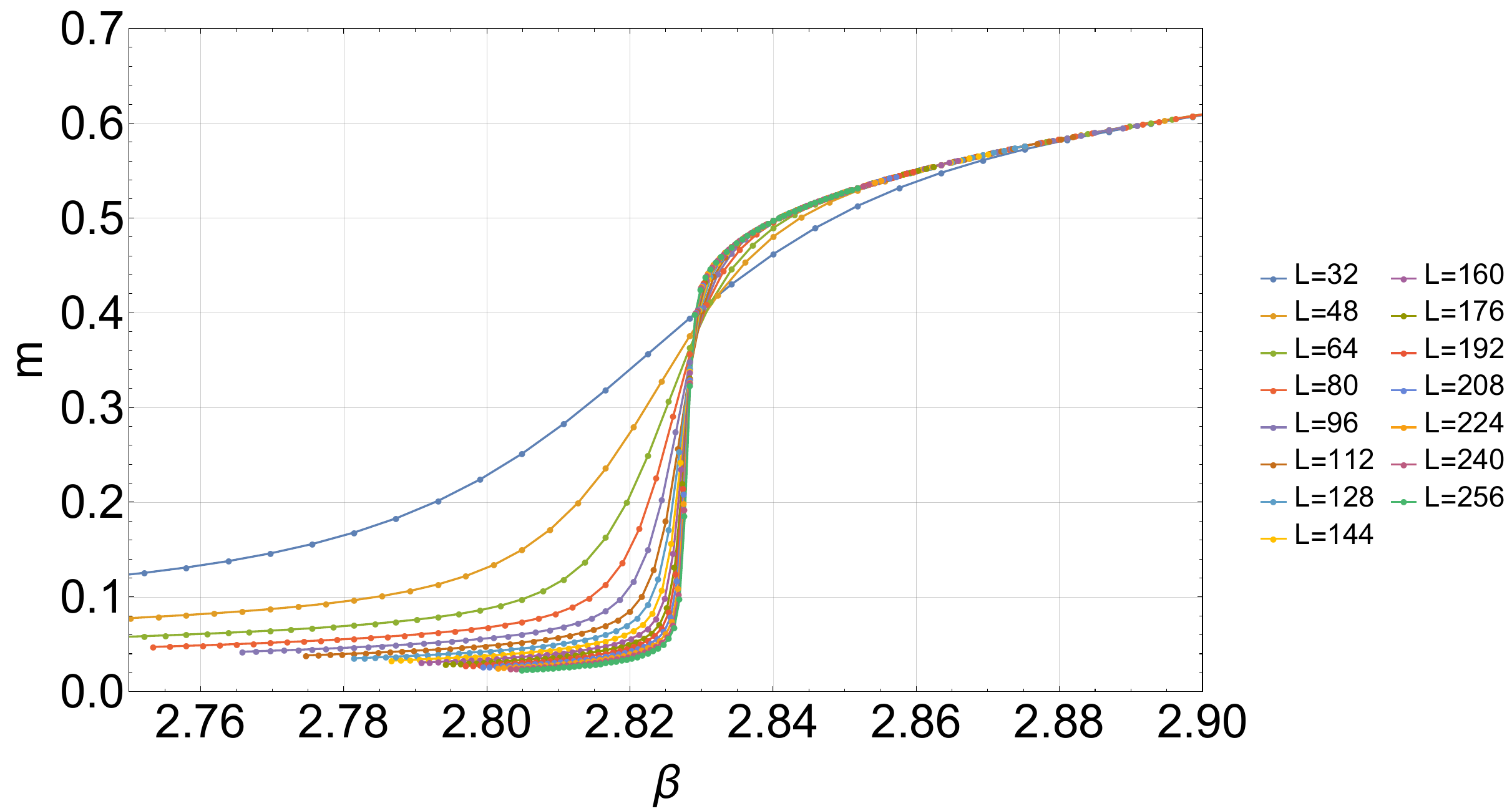}
    \caption{Magnetisation as a function of inverse temperature $\beta$ of a $q=3$ 2D Potts model with $z=88$ interacting neighbours in a narrow region around the pseudo critical point. 
    Each line represents a different system size in a range from $L_{min}=32$ to $L_{max}=256$. As the system size increases, the curve becomes steeper.}
    \label{fig_m_T}
\end{figure}

We start by analysing the standard physical observables, i.e. magnetisation, magnetic susceptibility and specific heat. 
In Fig. \ref{fig_m_T}, a magnetisation dependency on inverse temperature is shown for the case $z=88$ near the pseudo critical point for system sizes ranging from $L=32$ to $L=256$. 
As the system size increases, the dependency becomes steeper. However, since we are dealing with finite-size systems, we won't be able to observe a jump in the order parameter.
\begin{figure}
    \centering
    \includegraphics[width=0.46\textwidth]{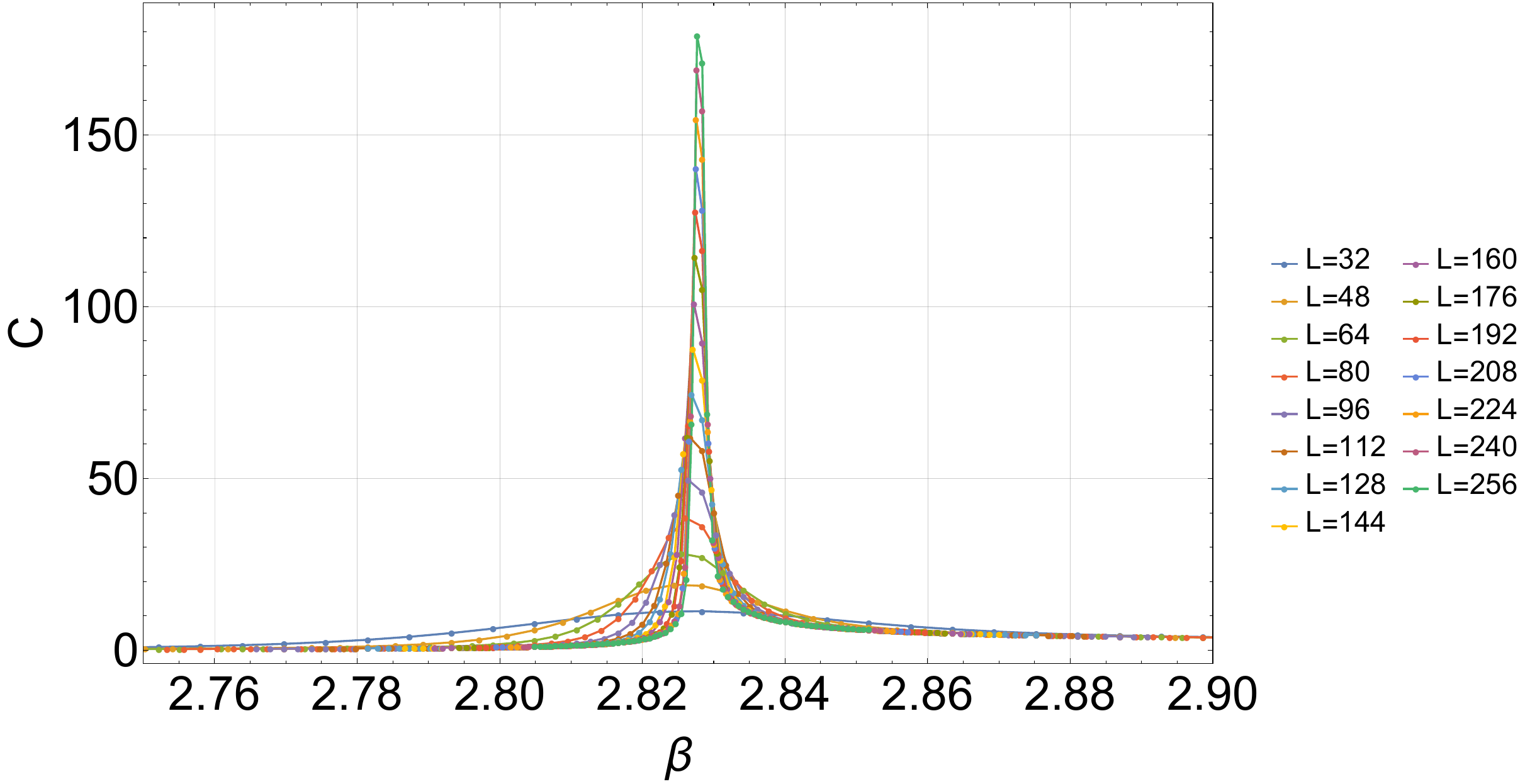}
    \includegraphics[width=0.46\textwidth]{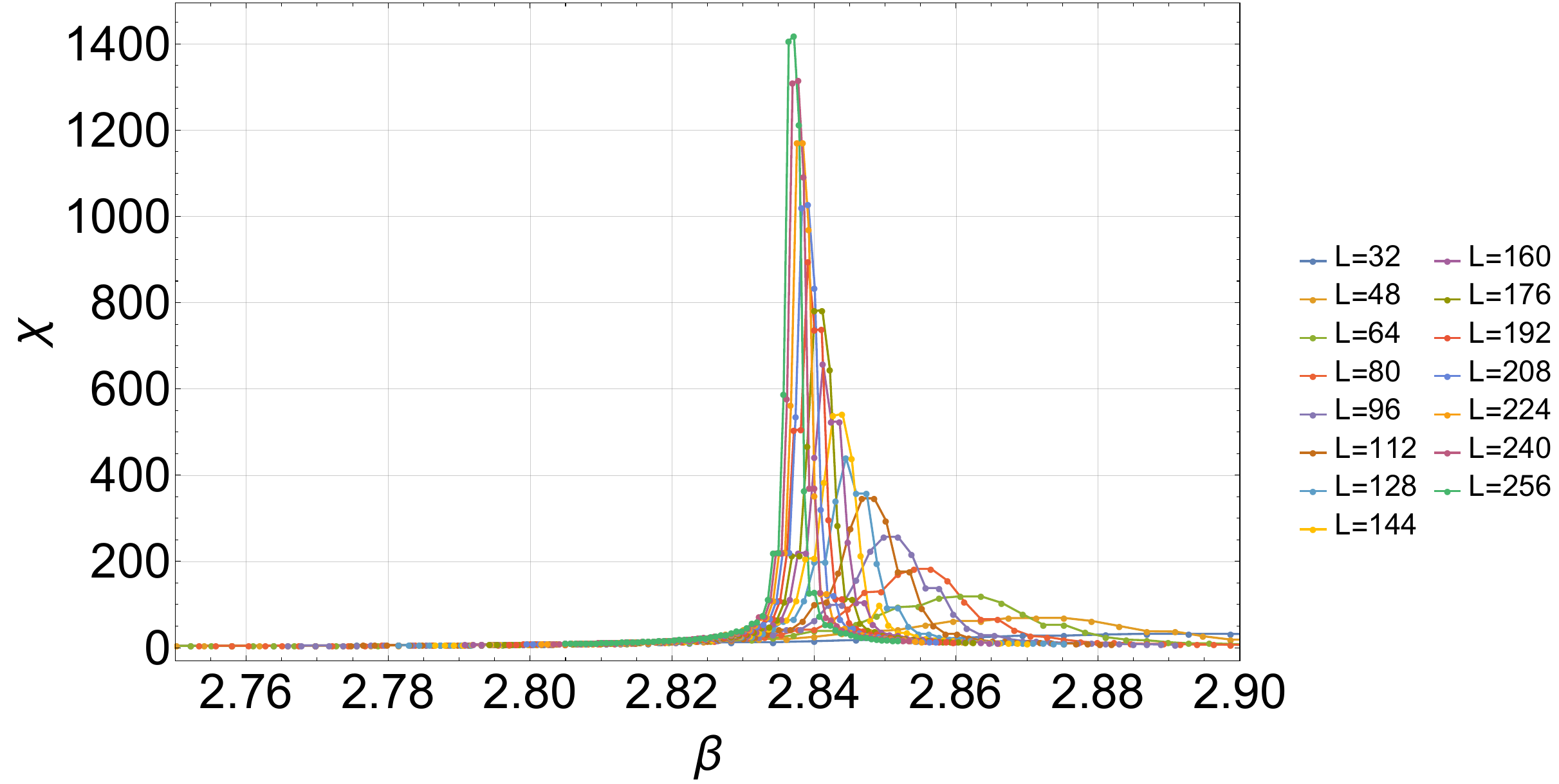}\\
    \hspace{4cm}(a)\hfill(b)\hspace{4cm}
    \caption{Specific heat (a) and magnetic susceptibility (b) as functions of inverse temperature $\beta$ of a $q=3$ 2D Potts model with $z=88$ interacting neighbours near the critical point. 
    Each line represents a different system size in a range from $L_{min}=32$ to $L_{max}=256$. As the system size increases, the peaks of both observables get higher and more narrow.}
    \label{fig_c_T}
\end{figure}
Additionally, in Fig. \ref{fig_c_T} specific heat and magnetic susceptibility are shown as a function of inverse temperature for the same case $z=88$. As expected, the peaks are getting higher and more narrow as the system size increases. The heights of the peaks scale with the system size according to the formulas
\begin{gather}
    C_{max}\propto L^{\alpha/\nu}\quad,\\
    \chi_{max}\propto L^{\gamma/\nu}\quad.
\end{gather}
Fitting the data to this ansatz yields $\alpha/\nu=1.334(4)$ and $\gamma/\nu=1.853(1)$. For the first-order regime, the scaling exponents are expected to be equal to the dimensionality $d=2$. Therefore, we will shift to the analysis of partition function zeros to obtain more accurate values of critical exponents.

\begin{figure}
    \centering
    $z=68$\\
    \includegraphics[height=0.42\textwidth]{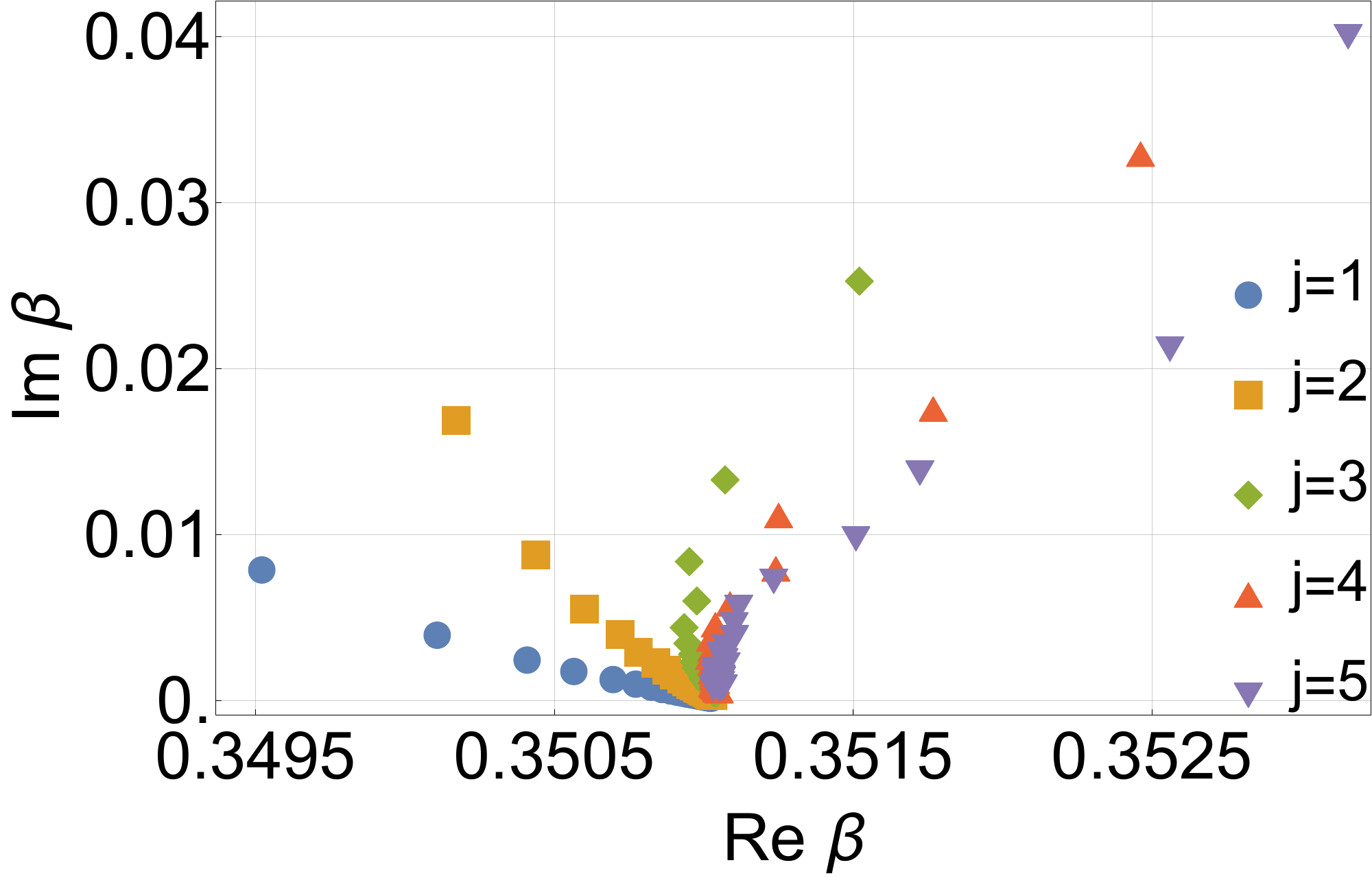}
    \caption{Coordinates of the first five Fisher zeros for the model with $z=68$. Different colours represent different sequential orders of a zero $j$. For a fixed $j$, the larger the system size $L$, the closer the zero lies towards the critical point.}
    \label{figz68_zeros}
\end{figure}

Using the method described in Section \ref{II}, we were able to calculate the coordinates of the first five Fisher zeros for each combination of the system size and the number of interacting neighbours. In Fig. \ref{figz68_zeros} we show the location of the zeros for the case $z=68$ and system sizes varying from $L_{\rm min}=32$ up to $L_{\rm max}=432$. Colours represent different sequential order of zeros: first, second, etc..

\begin{figure}
    \centering
    $z=68$\\
    \includegraphics[height=0.42\textwidth]{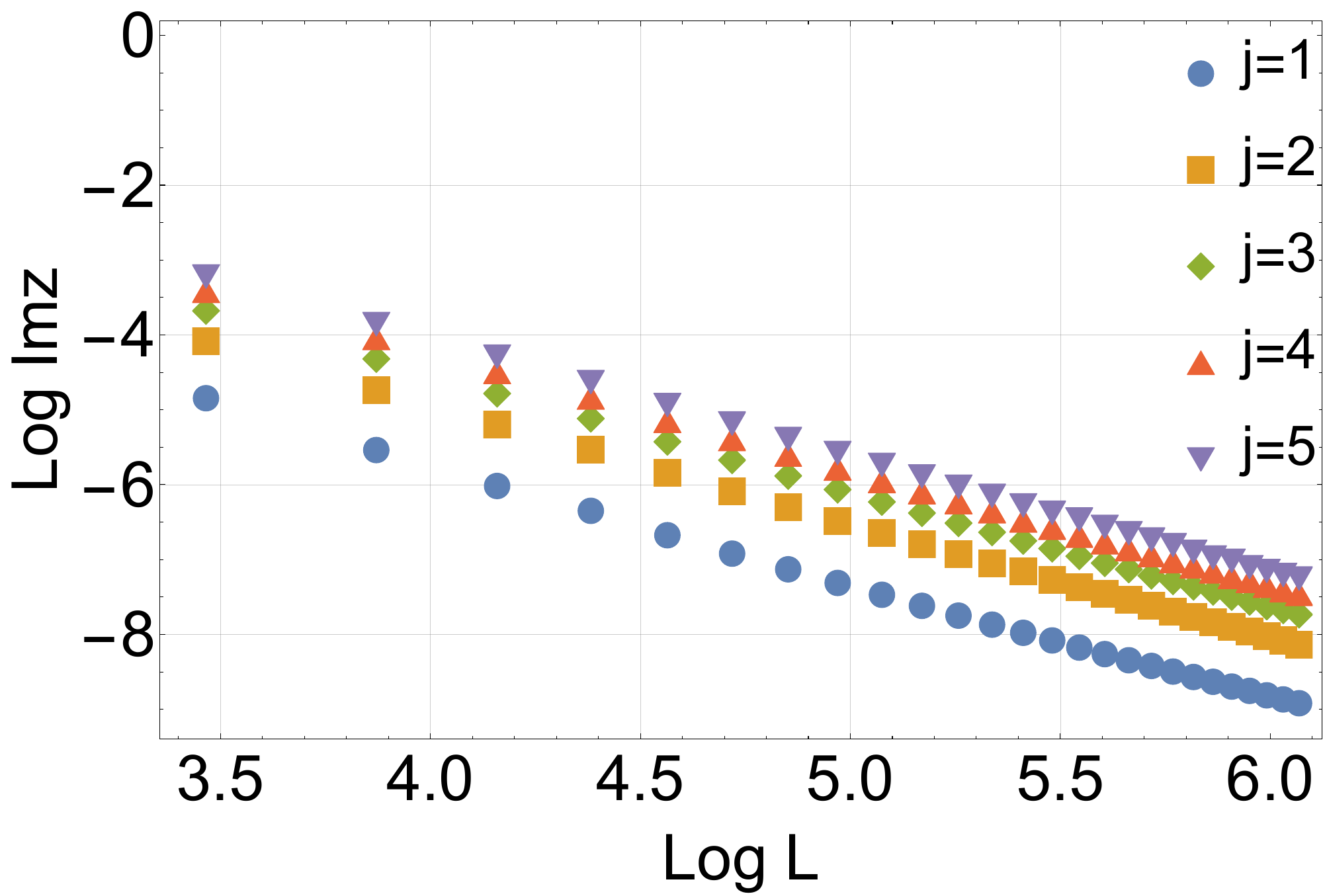}\\
    (a)\\
    \includegraphics[height=0.42\textwidth]{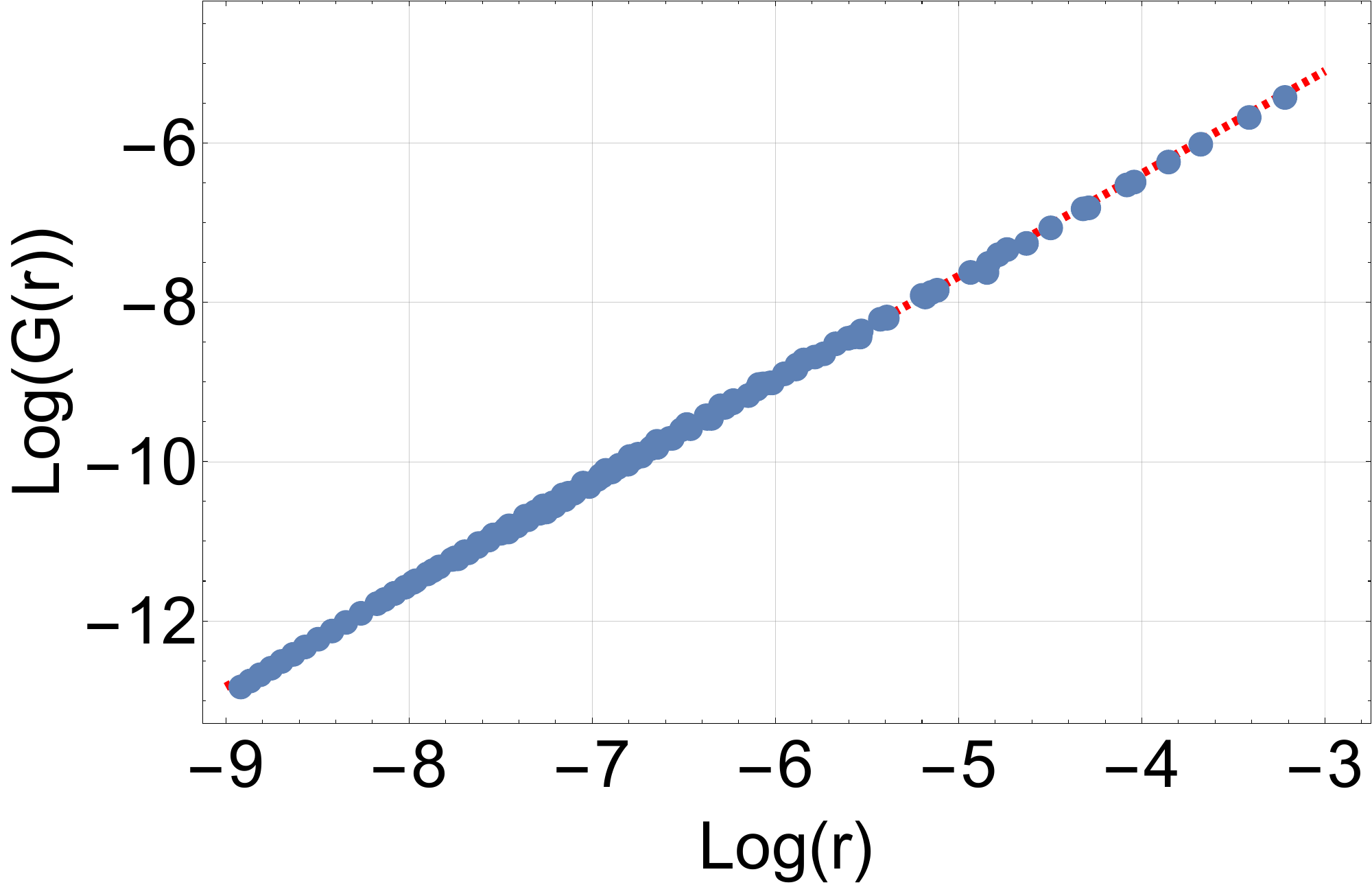}\\
    (b)
    \caption{(a): dependency of the imaginary part of Fisher zeros on the logarithm of a system size. Different colours represent different sequential numbers of zeros as shown in the legend. The closest to the real axis zero is shown in blue. The dependency is qualitatively as expected from Eq. (\ref{ansatz}). (b): logarithm of the partition function zeros density and its fit according to Eq. (\ref{density_scaling}). }
    \label{figz68_1}
\end{figure}

As it was mentioned in the previous section, one can analyse the finite-size scaling of the first, closest to the real axis, zero. 
According to the ansatz (\ref{ansatz}), this allows to extract the correlation length critical exponent $\nu$. 
Figure~\ref{figz68_1}(a) shows the dependence of the imaginary part of zeros on the logarithm of the system size for $z = 68$, as a representative example. 
As expected from Eq.~(\ref{ansatz}), the data points form an approximately straight line. 
A linear fit yields the critical exponent value $\nu=0.64148(2)$. 
The high precision of this estimate only reflects the excellent alignment of the data points along the fitted line as the expected critical exponent value is $\nu=5/6\simeq0.83333$.

Alternatively, the partition function zero density analysis takes into account not only the first closest zero, but also all the available ones. 
Effectively, this allows for enriching the data without the need to do additional simulations. 
In Fig. \ref{figz68_1}(b), the density of partition function zeros and its fit with a power law are given for case $z=68$. 
The corresponding specific heat critical exponent is calculated as $\alpha=0.71125(2)$.

\begin{figure}
    \centering
    $z=68$\\
    \includegraphics[height=0.42\textwidth]{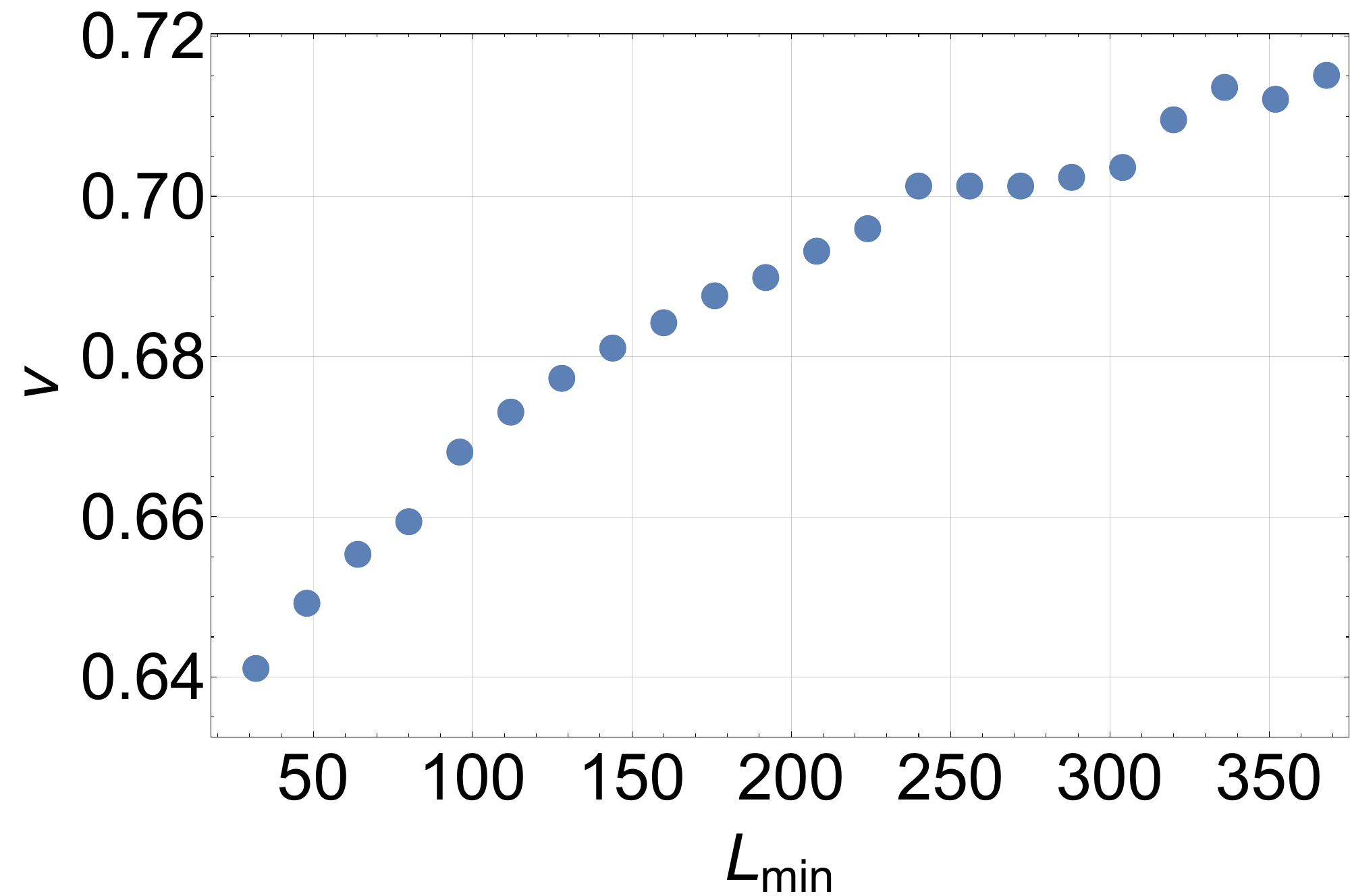}\\
    (a)\\
    \includegraphics[height=0.42\textwidth]{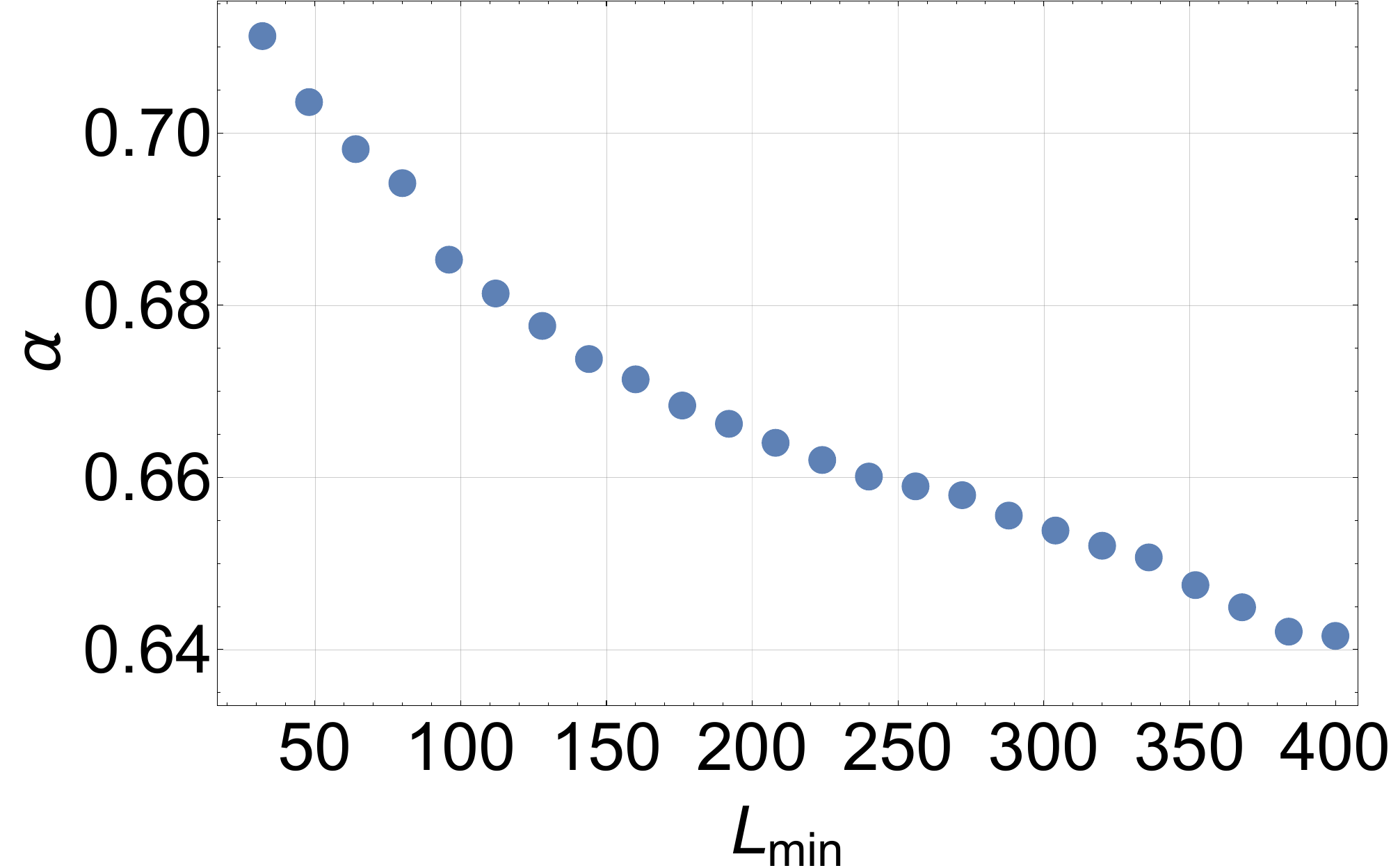}\\
    (b)
    \caption{Critical exponents values obtained from the partition function zeros analysis. As expected, when the fit is based on larger system sizes only, the result is shifted closer to the expected value. (a): correlation length critical exponent $\nu$ as a function of the minimal system size taken into consideration, obtained from the scaling of the closest zero.
    (b): specific heat critical exponent $\alpha$ as a function of the minimal system size taken into consideration, obtained from the zeros density analysis.}
    \label{figz68_2}
\end{figure}

The critical exponents obtained deviate significantly from the expected values. One possible explanation for this discrepancy is the presence of scaling corrections. 
This can be tested by excluding smaller system sizes from the analysis, as they tend to introduce greater statistical noise. 
Figure~\ref{figz68_2} shows the values of the critical exponent as a function of the smallest system size included in the fit, for the case of $z = 68$. 
It is evident that removing smaller system sizes shifts the estimated critical exponent closer to the value expected for the nearest-neighbour case. 
This observation suggests that scaling corrections play a substantial role in the model under investigation.

Performing a similar procedure for all the interaction ranges, one can obtain the critical exponents given in Table \ref{tab2}. 
For low values of $z$ we observe that small system sizes tend to shift the exponent $\nu$ to the lower values, while for large values of $z$ the trend is opposite. 
Similar behaviour is also present for $\alpha$, but since the two critical exponents are connected through the hyperscaling relation $\alpha+d\nu=2$ the trend with change in $z$ is opposite to that for $\nu$.
\begin{table}[h]
    \centering
    \begin{tabular}{|c|c|c|c|c|c|c|}
    \hline
    $z$ &  $\nu$ & $\alpha$ &  $\nu_{max}$ & $\alpha_{max}$ &  $\nu_{\rm extr}$ & $\alpha_{\rm extr}$ \\
    \hline
    68 &  $0.64106(2)$ & $0.71125(2)$ &  $0.7151(8)$ & $0.6416(2)$ & $0.822(6)$ & $0.342(24)$\\
    \hline
    72 &  $0.63290(2)$ & $0.73011(2)$ &  $0.6919(4)$ & $0.69553(8)$ & $0.811(6)$ & $0.471(15)$\\
    \hline
    76 &  $0.61059(2)$ & $0.76708(1)$ &  $0.6534(3)$ & $0.75228(6)$ & $0.798(23)$ & $0.591(22)$\\
    \hline
    80 & $0.59106(2)$ & $0.801884(8)$ &  $0.6212(2)$ & $0.80646(5)$ & $0.726(10)$ & $0.811(13)$\\
    \hline
    84 & $0.59289(1)$ & $0.813233(8)$ &  $0.6075(2)$ & $0.83841(4)$ &$0.614(3)$ & $0.891(7)$\\
    \hline
    88 & $0.56805(2)$ & $0.840474(9)$ &  $0.5664(2)$ & $0.86593(5)$ & $0.52(4)$ & $1.01(2)$\\
    \hline
    92 & $0.566094(3)$ & $0.83830(3)$ &  $0.5536(9)$ & $0.8555(2)$ & $0.52(3)$ & $0.94(2)$\\
    \hline
    96 & $0.55793(3)$ & $0.84823(3)$ &  $0.5249(9)$ & $0.86723(9)$ & $0.521(8)$ & $0.969(6)$\\
    \hline
    100 & $0.53871(1)$ & $0.910917(5)$ &  $0.52141(7)$ & $0.94869(2)$ & $0.499(2)$ & $0.99(1)$\\
    \hline
\end{tabular}
  
    \caption{Values of the correlation length critical exponent $\nu$ and of the specific heat critical exponent $\alpha$ obtained using the partition function zeros analysis. The second and the third columns are obtained using all the simulation data, while the fourth and the fifth columns were obtained using the data for the three largest system sizes. The sixth and seventh columns show the values of critical exponents obtained from the available simulation data and extrapolating the system size to infinity.}
    \label{tab2}
\end{table}
\section{Discussion}
\label{IV}

In Section \ref{III}, we focus on the particular value $z=68$; however, simulations were also performed for other interaction ranges mentioned in Table \ref{tab1}.
In each of these cases, we calculated the correlation length critical exponent $\nu$ and the specific heat critical exponent $\alpha$ to compare them with different scaling regimes shown in Table \ref{tab0} and, therefore, distinguish between the first- and second-order transitions.
In the nearest neighbours limit, when $z=4$, the $q=3$ 2D Potts model undergoes a second-order phase transition with critical exponents $\alpha=1/3\approx0.333$ and $\nu=5/6\approx0.833$ \cite{Wu1982}.
On the other hand, in the mean-field limit $z\to\infty$ the $q=3$ 2D Potts model undergoes a first-order phase transition, where the computed values of scaling exponents are $\alpha=1$ and $\nu=1/2$ \cite{Janke2002} \footnote{We use the same notation as for the second-order regime, even though the transition is of a first order.}.
If there is a tricritical point, the expected values of the critical exponents are $\alpha=5/6\approx0.833$ and $\nu=7/12\approx0.583$ \cite{nienhuis1987coulomb}.
\begin{figure}
    \centering
    \includegraphics[height=0.42\textwidth]{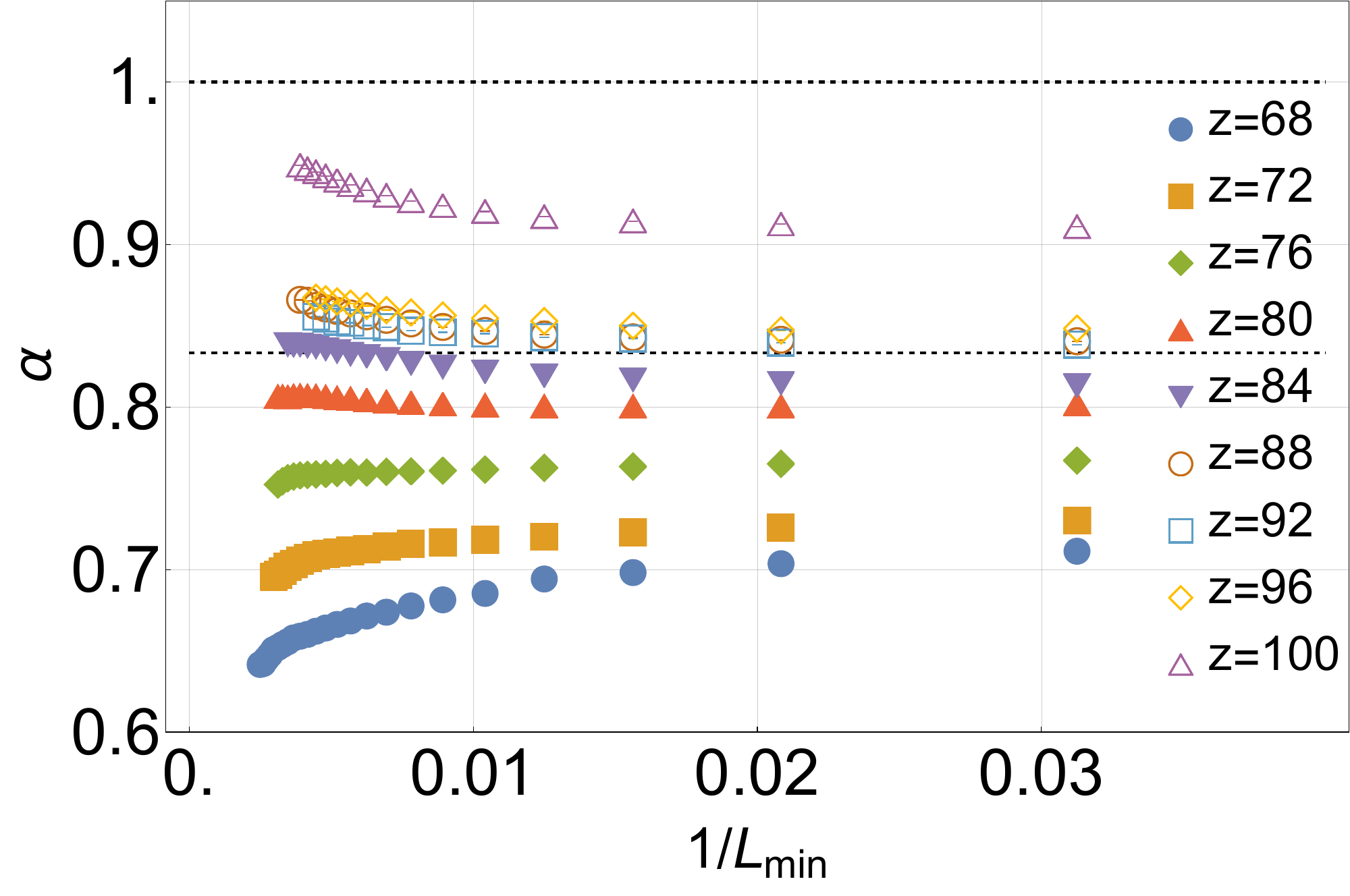}\\
    (a)\\
    \includegraphics[height=0.42\textwidth]{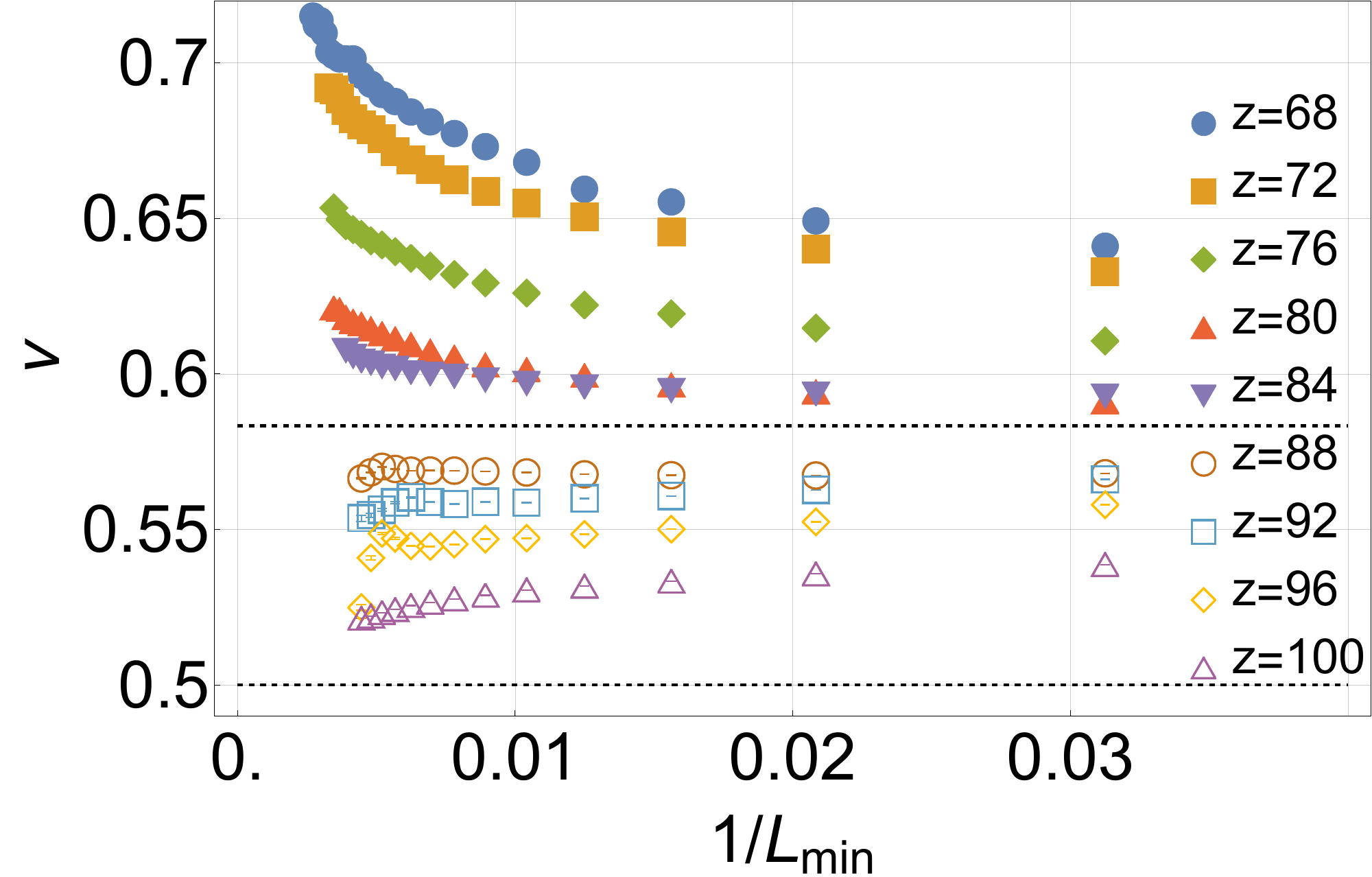}\\
    (b)\\
    \caption{Critical exponent values obtained from the partition function zeros analysis for different interaction ranges as a function of the inverse minimal system size taken into consideration. (a): specific heat critical exponent $\alpha$, the upper dashed line represents the first-order transition value, while the lower one represents the tricritical regime. (b): correlation length critical exponent $\nu$, the lower dashed line represents the first-order regime value, while the upper one is for the tricritical value.}
    \label{fig_all_z}
\end{figure}

In Fig. \ref{fig_all_z} we summarise the values of critical exponents obtained for different interaction ranges. 
Fig. \ref{fig_all_z}(a) shows the value of the specific heat critical exponent $\alpha$  for different $z$. 
The upper dashed line correspond to the $\alpha=1$, which is observed for the first-order transition.
While the lower dashed line represent $\alpha=5/6$ expected for the tricritical point. 
It is evident that corrections to the scaling have a significant effect on the estimated critical exponent for each interaction range.
However, for different interaction ranges, the corrections seem to have a different influence on the critical exponent. 
For $z<80$, with the increase of the system size the estimated value of $\alpha$ decreases. 
On the other hand, for $z\geq 80$ we see $\alpha$ increasing with $L_{min}$. 
However, what we observe is still an effective critical exponent of a finite-size system. In order to obtain these values in the thermodynamic limit, we extrapolate the obtained results. 
As it was mentioned in Ref. \cite{Bloete2016}, the corrections to the critical exponents are proportional to the $1/\log(L)$. We also used this ansatz to estimate the extrapolated $\nu_{\rm extr}$ and $\alpha_{\rm extr}$ values for the infinite system. In the case of $z=68$ we obtained $\nu_{\rm extr}=0.822(6)$ and $\alpha_{\rm extr}=0.342(24)$. These values are much closer to the expected values $\nu=5/6\approx0.8333$ and $\alpha=1/3\approx0.3333$. The values of extrapolated critical exponents for the other interaction ranges are given in the last two columns of Table \ref{tab2}.

It is worth noting here that for the nearest neighbours case $\alpha=1/3$ is only observed on the values obtained from extrapolation. This is another evidence that the corrections strongly affect the scaling.
For $z=80$, the specific heat critical exponent is close to the expected tricritical value. 
This suggests that the marginal value of the interaction range lies in this region.

In Fig. \ref{fig_all_z}(b), the values of the correlation length critical exponent $\nu$ are shown for the same set $68\leq z\leq 100$. 
The upper dashed line indicates the tricritical value, whereas the lower one is for the first-order regime.
The obtained critical exponents seem to strongly depend on the system sizes taken into consideration. 
Here for $z\geq88$ $\nu$ decreases when $L_{min}$ increases, while for all the other $z$ the behaviour is opposite. 
Moreover, extrapolated values for $z\geq 88$ are approaching the first-order regime in the thermodynamic limit. For $z<80$ the extrapolated values of the correlation length critical exponent are close to the nearest neighbours case $\nu=5/6$.

The two most challenging cases are $z=80$ and $z=84$. Both of these values seem to be in the crossover region. In the former case, the extrapolated specific-heat critical exponent is close to its tricritical value, whereas $\nu_{\rm extr}$ remains near the second-order regime. Considering the trends in the critical exponents shown in Fig.~\ref{fig_all_z}, we argue that $z=80$ still belongs to the second-order region. In contrast, for $z=84$, the correlation-length critical exponent approaches its tricritical value, while the specific-heat critical exponent tends toward the first-order value. Based on the observed trends in Fig.~\ref{fig_all_z}, we conclude that this case already lies within the first-order regime. Therefore, the marginal value of $z$ separating the regions with distinct critical behaviour appears to fall within the range $z=80\div84$.

We employed two different approaches to analyse the partition function zeros. 
From the scaling behaviour of the zero closest to the real axis and the density of Fisher zeros, we inferred that the marginal value lies in the range $z = 80 \div 84$. 
This result is consistent with the previously estimated value $z_c = 80$ reported in Ref.~\cite{Bloete2016}. 
A more precise determination of $z_c$ could be achieved by increasing statistics, i.e., through longer simulation runs. 
Nonetheless, $z$ cannot be set to arbitrary values, as doing so would break the symmetry of the interaction.

\vspace{6pt}

\begin{acknowledgments}
I am deeply grateful to the late Professor Ralph Kenna, who first inspired and stood firmly behind the idea of this research. His unwavering enthusiasm, intellectual curiosity, and boundless energy created an atmosphere that encouraged exploration and creativity. His influence continues to guide this work and all of us who had the privilege to learn from him. I would also like to thank Professor Yurij Holovatch for fruitful discussions and comments. This research was funded by the National Research Foundation of
Ukraine, Project 2023.03/0099 “Criticality of complex
systems: fundamental aspects and applications".
\end{acknowledgments}

\bibliography{references.bib}

\end{document}